\documentclass[pra,twocolumn,longbibliography]{revtex4-2}
\usepackage[colorlinks,urlcolor=blue,citecolor=blue,linkcolor=blue]{hyperref}

\usepackage{graphicx}
\usepackage{amsmath}
\usepackage{amssymb}
\usepackage{bm}
\usepackage{mathtools}
\usepackage{graphics}
\usepackage{comment}
\usepackage{tikz}
\usepackage{mathtools}
\usepackage{physics}

\renewcommand{\v}{\mathbf{v}}
\newcommand{\f}{\mathbf{f}}
\newcommand{\q}{\mathbf{q}}
\newcommand{\Q}{\mathbf{Q}}
\renewcommand{\O}{\mathbf{\Omega}}
\newcommand{\e}{\bm{\epsilon}}
\newcommand{\ez}{\mathbf{e}_z}
\newcommand{\er}{\mathbf{e}_r}
\newcommand{\Jp}{J_{v+}}
\newcommand{\Jm}{J_{v-}}

\newcommand{\cT}{c_\mathrm{T}}
\newcommand{\V}{\mathbf{V}}
\renewcommand{\u}{\mathbf{u}}
\newcommand{\s}{\mathfrak{s}}
\newcommand{\st}{\tilde{\mathfrak{s}}}
\newcommand{\A}{\mathbf{A}}

\newcommand{\Jepsp}{J_{\epsilon+}}
\newcommand{\Jepsm}{J_{\epsilon-}}

\newcommand{\Ylm}{\mathbf{Y}_{lm}}
\newcommand{\Psilm}{\mathbf{\Psi}_{lm}}
\newcommand{\Philm}{\mathbf{\Phi}_{lm}}

\newcommand{\lm}{{lm}}

\newcommand{\sch}{Schr{\"o}dinger}

\newcommand{\diff}{\mathop{}\!\mathrm{d}}

\newcommand{\nn}{\nonumber}

\renewcommand{\vec}{\mathbf}

\renewcommand{\vec}{\mathbf}

\begin{document}

\title{Equatorial Waves in Rotating Bubble-Trapped Superfluids}

\author{Guangyao Li}
\affiliation{School of Physics and Astronomy, Monash University, Victoria 3800, Australia}
\affiliation{ARC Centre of Excellence in Future Low-Energy Electronics Technologies, Monash University, Victoria 3800, Australia}

\author{Dmitry K. Efimkin}
\affiliation{School of Physics and Astronomy, Monash University, Victoria 3800, Australia}
\affiliation{ARC Centre of Excellence in Future Low-Energy Electronics Technologies, Monash University, Victoria 3800, Australia}

\begin{abstract}
As the Earth rotates, the Coriolis force causes various oceanic and atmospheric waves to be trapped along the equator, including Kelvin, Yanai, Rossby, and Poincar\'{e} modes. It has been demonstrated that the mathematical origin of these waves is related to the nontrivial topology of the underlying hydrodynamic equations. Inspired by recent observations of Bose-Einstein condensation (BEC) in bubble-shaped traps in microgravity ultracold quantum gas experiments, we demonstrate that equatorial modes are supported by a rapidly rotating condensate in a spherical geometry. Using a zero-temperature coarse-grained hydrodynamic framework, we reformulate the coupled oscillations of the superfluid and the Abrikosov vortex lattice resulting from rotation as a \sch-like eigenvalue problem. The resulting non-Hermitian Hamiltonian is topologically nontrivial. We also solve the hydrodynamic equations for a spherical geometry and find that the rotating superfluid hosts Kelvin, Yanai, and Poincar\'{e} equatorial modes, but not the Rossby mode. Our predictions can be tested with state-of-the-art bubble-shaped trapped BEC experiments.
\end{abstract}

\maketitle

\section{Introduction}

The Earth's rotation causes a Coriolis force in the dynamics on its surface, which in turn deflects ocean currents, shapes cyclones, and plays a central role in the climate. The reversal of the Coriolis effect across the equator traps some ocean and atmospheric waves, with Kelvin and Yanai modes propagating only toward the east. Although scientists have long known about these modes \cite{REV_FEDOROV2009271}, the chiral nature and exceptional robustness of these modes have recently been shown to be of a topological nature~\cite{DelplaceScience2017}. These modes are similar to protected states in topological states of matter~\cite{TopReview1,TopReview2}.    

We extend the above idea by considering what would happen to these equatorial waves if the ocean were a superfluid. This type of scenario occurs in neutron star dynamics  \cite{Book_neutron_star}.
While it is difficult to conduct experiments in such an exotic environment, we can address this question through recent reports on bubble-trapped Bose–Einstein condensates (BECs) \cite{CarolloNature2022}. Because the density of a BEC in a shell-shaped profile is very sensitive to the anisotropy caused by Earth's gravitational field, it is necessary to engineer artificial microgravity setups to allow atoms to uniformly cover a bubble’s entire surface. Several schemes for facilitating the required microgravity environment are readily available, including free-fall drop tower \cite{Zoest_Science_2010}, zero-G aircraft \cite{Barrett_NatComm_2016},  Einstein elevator \cite{Condon_PRL_2019}, and planetary orbit \cite{Elliott_npjMicrogravity2018, AvelineNature2020,CarolloNature2022} setups.
In principle, each of these schemes is capable of generating shell-shaped BECs.  Experimental advances have led to a plethora of theories focusing on non-flat geometries, e.g., bubble-trap inflation and manifold curvature effects \cite{SunPRA2018,TononiPRL2020,RhynoPRA2021,TononiPRA2022,WolfPRA2022}, dynamics of individual vortices and thermodynamics of vortex clusters \cite{FetterPRA2021,FetterPRA2022,VortexAntiVortex_PRA2020}, quantum scattering and interactions \cite{TononiPRL2019,PrestipinoPRA2019,Arazo_2021}, and the Berezinskii–Kosterlitz–Thouless superfluid transition \cite{TononiPRR2022}.

\begin{figure}[b]
	\centering
	\vspace{-0.1in}
	\includegraphics[width=.8\linewidth]{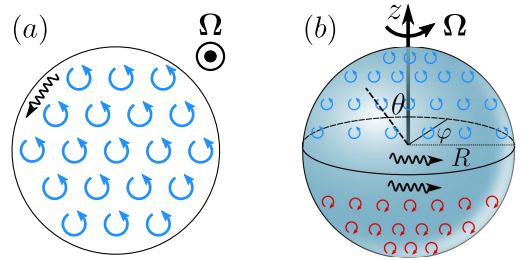}
	\caption{ (a) A rotating superfluid in a planar configuration hosts a single chiral edge mode. (b) A bubble-trapped superﬂuid supports a pair of chiral states that propagate along the equator of the trap. These states are analogous to equatorial ocean Kelvin and Yanai waves trapped by the Earth's rotation. These chiral states can be detected and studied using experimental techniques such as two-photon Bragg scattering and dynamic structure factor measurements. }
	\label{fig:sphere_illustration}
	\vspace{-0.2in}
\end{figure}

In this work, we re-examine the coarse-grained hydrodynamic equations describing a rotating superfluid and the Abrikosov vortex lattice within it. When the geometry is planar, these equations can be transformed into a matrix eigenvalue problem. The resulting matrix functions as a non-Hermitian Hamiltonian and belongs to the $D^\dagger$ class \cite{KawabataPRX2019,AshidaAdvanPhys2020}. It also exhibits nontrivial topological properties.
As a result, we predict the existence of a single chiral mode that moves around a uniform, rotating superfluid in a flat-bottomed optical box trap~\cite{FlatTrapReview1}, as shown in Fig.~\ref{fig:sphere_illustration}(a).
In the spherical geometry shown in Fig.~\ref{fig:sphere_illustration}(b), the rotating superfluid may contain chiral Kelvin and Yanai modes that are similar to the equatorial waves trapped by the Earth. However, the topological protection of these chiral edge modes is limited, as they can decay into Tkachenko shear sound waves supported by the Abrikosov lattice over a wide range of frequencies.
Our theory offers a novel approach to probe the unconventional rotation of superfluids by exploring the interactions between the vortex lattice's vibrational modes and the equatorial modes. By utilizing advanced rotating bubble-trapped superfluid experimental setups, we can investigate the interplay between topology, manifold curvature, and vortex physics in superfluids, which are fundamental concepts of great significance.

\section{The hydrodynamic framework}

It is instructive to start with a discussion of the dynamics of a rapidly rotating, uniform BEC with a planar geometry. At low temperatures, the normal component can be safely neglected. The dynamics of the superfluid and the Abrikosov vortex lattice in a co-rotating frame can be described by the following linearized coarse-grained hydrodynamic equations~\cite{BaymPRL2003,SoninPRA2005,Baym1986,SoninRMP1987}: 
\begin{align}
   & \partial_t \rho +\rho_0\nabla \cdot \v = 0, \label{eq:hydro_plane_1}   \\
   & \partial_t \v= -2 \O\times \v-\frac{c^2}{\rho_0}\nabla \rho +\f, \label{eq:hydro_plane_2}\\
   & 2\O\times\left(\partial_t\e-\v\right)=-\f.\label{eq:hydro_plane_3}
\end{align}
Here, $\rho$ represents the fluctuation in superfluid density above the equilibrium density $\rho_0$, $\v$ is the superfluid velocity, $\e$ is the vortex displacement from its equilibrium position, $\O=\Omega\, \ez$ is the superfluid's rotation angular velocity normal to the plane, and $c$ is the speed of sound in the condensate
\cite{BookPitaevskii}. Eq.~\eqref{eq:hydro_plane_1} is the continuity equation for the conservation of mass; Eq.~\eqref{eq:hydro_plane_2} is Euler's equation for a superfluid, where the terms on the right-hand side representing the Coriolis force, pressure force, and vortex lattice elastic force~\footnote{We assume that the centrifugal force is negligibly small or compensated by the trap potential}. The last equation describes the vortex displacement in response to the elastic force density: 
\begin{equation}
\label{eq:force}\f=c_{\mathrm{T}}^2 \left[\nabla^2 \e  -\alpha\,2\nabla(\nabla\cdot\e)  \right],
\end{equation}
where the first term inside the square bracket comes from the shear stress of the Abrikosov vortex lattice and the second term comes from the compressional stress. The magnitudes of these terms can be parameterized by the Tkachenko wave velocity $\cT$~\cite{BaymPRL2003,SoninPRA2005,Baym1986,SoninRMP1987} and a dimensionless parameter $\alpha$, which determines their relative strength. Microscopic calculations for the Abrikosov lattice give values of $\cT=\sqrt{\abs{\Omega}/4m_a}$ and $\alpha=1$, with $m_a$ being the atomic mass. However, the compressional contribution to $\f$ is frequently neglected (i.e., $\alpha=0$) in previous studies \cite{BaymPRL2003,SoninPRA2005}. By treating $\alpha$ as a free parameter that ranges from $0$ to $1$, we can continuously adjust the system's dynamics from an approximated one to the full dynamics.
In our numerical calculations, we use characteristic energy and momentum units of $\mathcal{E}_0=m_a c^2$ and $p_0=m_a c$, and we assume $\hbar=1$ in most of the following discussions.

Remarkably, Eqs.~(\ref{eq:hydro_plane_1})–(\ref{eq:hydro_plane_3}) can be rewritten as a Schr\"{o}dinger-like eigenvalue problem:
$\omega \psi=\hat{\mathcal{H}}(\q,\alpha) \psi$, where $\psi\equiv\psi(\q,\omega)$. 
To do this, we perform a Fourier transform and introduce the spinor $\psi=[\Jepsp,\Jp,J_0,\Jm,\Jepsm]^\mathsf{T}$, where $J_{v\pm}\equiv \rho_0(v_x\pm i v_y)/\sqrt{2}$, $J_{\epsilon\pm}\equiv \rho_0 \cT q (\epsilon_x\pm i \epsilon_y)/\sqrt{2}$, and $J_0\equiv c \rho$.  The resulting $5\times 5$ matrix, $\hat{\mathcal{H}}(\q,\alpha)$, acts as a Hamiltonian and is equal to:
\begin{equation}\label{eq:H_alpha}
\hat{\mathcal{H}}(\q,\alpha)=
    \begin{bmatrix}
    \frac{ \bar{\alpha}c_{\mathrm{T}}^2 \q^2}{2\Omega} & i c_{\mathrm{T}} q   &    0    & 0 & -\frac{ \alpha c_{\mathrm{T}}^2 q^2_+ }{2\Omega}\\
    - i\bar{\alpha} c_{\mathrm{T}} q                  &   2\Omega  &   \frac{c q_+}{\sqrt{2}}  &  0  & i \frac{ \alpha c_{\mathrm{T}} q_+^2}{q}\\
    0       &   \frac{c q_- }{\sqrt{2}}  &   0   &      \frac{c q_+ }{\sqrt{2}}  &       0 \\
    i  \frac{ \alpha c_{\mathrm{T}} q_-^2 }{q}   &    0   &   \frac{c q_- }{\sqrt{2}}   &   -2\Omega    &   - i \bar{\alpha} c_{\mathrm{T}} q\\
     \frac{ \alpha c_{\mathrm{T}}^2 q_-^2}{2\Omega} &  0   &  0  &  i c_{\mathrm{T}} q &   - \frac{\bar{\alpha} c_{\mathrm{T}}^2 \q^2}{2\Omega}
    \end{bmatrix}.
\end{equation}
In the above equation, we have introduced $q_{\pm}=q_x\pm i q_y$ and $\bar{\alpha}=1-\alpha$. 
Unless $\alpha=0$, the Hamiltonian is non-Hermitian, and its complicated asymmetric expression does not suggests that its spectrum can be real.
However, this Hamiltonian follows the particle–hole symmetry       $\mathcal{S}\hat{\mathcal{H}}(\q,\alpha) \mathcal{S} ^{-1}=-\hat{\mathcal{H}}(-\q,\alpha)$, where $\mathcal{S}=\mathcal{A K}$ \cite{KawabataPRX2019}. Here, $\mathcal{A}$ is an anti-diagonal unit matrix, and $\mathcal{K}$ represents the complex conjugate. The presence of this particle–hole symmetry is closely connected to the classical nature of the problem and imposes strict restrictions on the spectrum. If the compressional contribution to the elastic force density is ignored (i.e., $\alpha=0$), the resulting Hamiltonian, $\hat{\mathcal{H}}_0(\q)\equiv\hat{\mathcal{H}}(\q,0)$, is a Hermitian matrix. 
\begin{figure}[t]
	\centering
	\includegraphics[width=\linewidth]{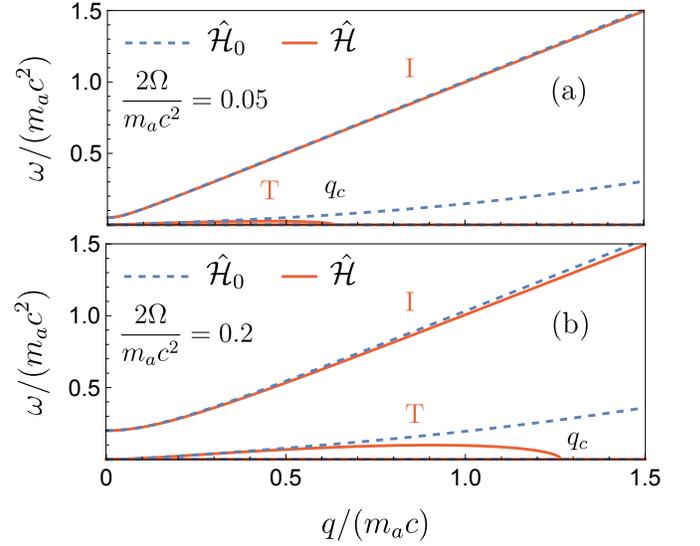}
	\caption{Dispersion relations for inertial (I) and Tkachenko shear sound (T) modes supported by a rotating superfluid in planar geometry. The dispersion relations were obtained using both the full Hamiltonian, denoted as $\hat{\mathcal{H}}(\q)$ (shown in solid red), and an approximate Hamiltonian, $\hat{\mathcal{H}}_0(\q)$ (shown in blue dashed), which neglects the compressional contribution to the elastic force density as described in Eq.~(\ref{eq:force}).
	}
	\label{fig:E5_dispersion}
	\vspace{-0.2in}
\end{figure}

The dispersion curves for modes supported by the rotating superfluid are given by positive-frequency eigenvalues for $\hat{\mathcal{H}}(\q)\equiv\hat{\mathcal{H}}(\q,\alpha=1)$~\footnote{The eigenvalues of $\hat{\mathcal{H}}(\q)$ include a single spurious zero-frequency mode, two positive frequency modes, and two negative frequency modes. However, the negative frequency modes are not dynamically independent from the positive frequency modes due to symmetries in the following discussions.}. In the characteristic units mentioned earlier, these modes depend only on the ratio $2|\Omega|/m_ac^2$ and are plotted in Fig.~2 for values of $0.05$ and $0.2$. The modes have two branches, and their long-wavelength behavior can be accurately approximated as follows:
\begin{equation}
\label{eq:Modes}
\omega_\mathrm{I}=\sqrt{c^2 \q^2 + (2\Omega)^2}, \quad \quad \omega_\mathrm{T}=\frac{c\, \cT \q^2}{\sqrt{c^2 \q^2 +(2\Omega)^2}}.    
\end{equation}
The first branch, $\omega_{\mathrm{I}}(\q)$, is the inertial mode (I) of the rotating superfluid and has a gap of $2|\Omega|$ in its dispersion relation. Essentially, it is the usual sound mode gapped by the rotation. The second branch, $\omega_{\mathrm{T}}(\q)$, is the Tkachenko sound mode (T), which represents coupled transverse oscillations of the vortex lattice and the superfluid \cite{BaymPRL2003,SoninJETP2014}. At long wavelengths both modes are $\alpha$-independent and can be accurately described by the Hermitian Hamiltonian $\hat{\mathcal{H}}_0(\q)$, which ignores the compressional contribution to the elastic force. However, the situation becomes more complex beyond the long-wavelength limit.

The dispersion of the Tkachenko mode predicted by $\hat{\mathcal{H}}_0(\q)$ increases with increasing value of $q$ and eventually exceeds the gap of $2|\Omega|$. In contrast, the growth of the Tkachenko mode predicted by $\hat{\mathcal{H}}(\q)$ reaches a maximum at an intermediate value of $q$ and then saturates. The mode becomes unstable at $q_{\mathrm{c}}=\sqrt{16 m_a|\Omega|}$, which is similar to the exceptional point found in parity–time symmetric systems \cite{PT_REV_2019}.
However, the value of $q_{\mathrm{c}}$ is comparable to size of the Brillouin zone edge for the Abrikosov lattice of vortices, $q_\mathrm{v}=\pi/l_\mathrm{v}$, where $l_\mathrm{v}=\sqrt{1/\sqrt{3}  m_\mathrm{a}|\Omega|}$
is the intervortex distance~\cite{Baym1983}.
This limits the validity of the hydrodynamic description, which requires $q\ll q_\mathrm{v}$. While the presence of an exceptional point in the spectrum of $\hat{\mathcal{H}}(\q)$ may seem to be an artifact of the hydrodynamic equations, the predicted saturation behavior of the Tkachenko mode is consistent with calculations based on the Bogoliubov–de Gennes equation~\cite{ TapioTkachenko}. 
The existance of a \emph{global} frequency gap separating the inertial and Tkachenko modes in the bulk spectrum is important for the complete protection of edge states, which are closely related to the nontrivial typology for $\hat{\mathcal{H}}(\q)$.

\section{Nontrivial topologies}

The Hamiltonian $\hat{\mathcal{H}}(\q)$ exhibits particle–hole symmetry, but it does not possess time-reversal symmetry, which is broken by the rotation. According to the recently developed classification of non-Hermitian matrices, $\hat{\mathcal{H}}(\q)$ belongs to the $\operatorname{D}^\dagger$ class \cite{KawabataPRX2019,AshidaAdvanPhys2020}, and according to the line-gap definition, each isolated dispersion curve is characterized by a topological Chern number, similar to a Hermitian system. As discussed in the Appendix, the Chern number for the inertial mode is equal to $\mathcal {C}=\Omega/|\Omega|$, but its value for the Tkachenko mode is not well defined~\footnote{At $\vec{q}=0$, the Tkachenko mode experiences band-touching not only with the spurious zero-frequency mode, but also with the particle–hole conjugate branch.}. Fortunately, the nontrivial topologies for the inertial mode can be easily tracked and do not depend on the mathematical complexities of non-Hermitian physics. 

One notable observation is that if $\alpha$ is varied continuously from $0$ to $1$, the dispersion curve for the inertial mode can smoothly transform without experiencing any band-touching. As a result, $\hat{\mathcal{H}}_0(\q)$ and $\hat{\mathcal{H}}(\q)$ are topologically equivalent \cite{KawabataPRX2019,AshidaAdvanPhys2020}. If we further remove the shear component in the elastic force density $\vec{f}$ (i.e., if the dynamics of the vortex lattice is frozen), the Tkachenko mode becomes the spurious zero-frequency mode, but the dispersion curve for the inertial mode remains almost unchanged. In this case, the dynamics can be well described by the Hamiltonian $\hat{\mathcal{H}}_0^\prime(\q)$, which is given by the central block in $\hat{\mathcal{H}}(\q)$ as follows:
\begin{equation}
\hat{\mathcal{H}}^\prime_0(\q)=
    \begin{bmatrix}
2\Omega  &   \frac{c q_+}{\sqrt{2}}  &  0  \\  \frac{c q_- }{\sqrt{2}}  &   0   &      \frac{c q_+ }{\sqrt{2}} \\
    0   &   \frac{c q_-}{\sqrt{2}}   &   -2\Omega    
    \end{bmatrix},
\end{equation}
which is also topologically equivalent to $\hat{\mathcal{H}}(\q)$. An analysis of topologies of $\hat{\mathcal{H}}^\prime_0(\q)$ is simple, and the resulting Chern number for the inertial mode is equal to $\mathcal {C}=\Omega/|\Omega|$, as previously mentioned. It is interesting to note that this Hamiltonian $\hat{\mathcal{H}}^\prime_0(\q)$ is the same as the one used in Ref.~\cite{DelplaceScience2017} to describe ocean equatorial waves on the surface of the Earth. This is not surprising, as the zero-temperature dynamics of a superfluid with a coarse-grained, continuous (or diffused) vorticity and a frozen Abrikosov lattice are similar to the hydrodynamics of an inviscid classical fluid. 

In a planar geometry, the bulk-edge correspondence suggests the presence of a single chiral inertial mode circulating around a uniform superfluid~\footnote{For spatially nonuniform superfluids in harmonic traps, the spectrum of edge states is more complicated but still has some chirality~\cite{TercasNJP_2010,EdgePlanar1,EdgePlanar2}. Such spectra have not been linked to nontrivial topologies for the superfluid dynamics revealed in this work.} in a flat-bottomed optical box trap~\cite{FlatTrapReview1,SergejPRB_1,SergejPRB_2}. The presence of the in-gap Tkachenko shear sound mode requires additional care when using topological arguments. When the Abrikosov vortex lattice is frozen ($\e=0$), the evaluation of edge states is straightforward. These states have a dispersion of $\omega_\mathrm{e}(q)=c q$ and a penetration length of $l_{\mathrm{e}}=c/2|\Omega|$. The range of reciprocal space $q_{\mathrm{e}}=2|\Omega|/c$ in which edge states reside is much smaller than $q_\mathrm{v}$, allowing them to be accurately described by the hydrodynamic framework. While there is still a global gap (between $\Omega$ and $2\Omega$) in the spectrum in the presence of shear oscillations of the vortex lattice, edge states within this gap have complete topological protection. However, edge states with frequencies below $\Omega$ can decay into bulk Tkachenko sound waves through transitions requiring a large momentum transfer of $\Delta q\sim\sqrt{q q_\mathrm{c}}$. This scattering will not be efficient when the superfluid density is sufficiently uniform and there are no intentionally introduced impurities or defects.

\section{Dynamics on a spherical surface}

Simulations of the Gross–Pitaevskii equation have shown that triangular vortex and antivortex Abrikosov lattices can form at the poles of a rotating bubble-trapped superfluid if its radius $R$ is only a few times larger than the intervortex distance $l_\mathrm{v}$~\footnote{Angela White, private communication; unpublished}. In this study, we assume that $R$ is more than one order of magnitude larger than $l_\mathrm{v}$, allowing the Abrikosov vortex lattice to adjust to the latitude-dependent radial component of the angular frequency $\Omega_\vec{r}(\theta)=\Omega \cos(\theta)$. This allows us to extend the hydrodynamic equations to a spherical geometry and safely neglect terms proportional to the derivative of $\Omega_\vec{r}(\theta)$. By expanding the resulting equations over vector spherical harmonics \cite{VSHwiki,CooperPRB2022}, which are labeled by the non-negative integer value for angular momentum $l$ and its $z$-axis projection $m=-l, -l+1, \dotsc ,l$, we can solve them in the presence of uniform rotation, where axial symmetry is maintained and $m$ is conserved. However, after extending Eq.~\eqref{eq:hydro_plane_1}-\eqref{eq:hydro_plane_3} to a spherical geometry, the resulting equations cannot be reduced to an eigenvalue problem due to the non-differentiable behavior of $\cT$ across the equator (see Appendix). The spectrum calculation was performed on the Monash Advanced Research Computing Hybrid with only six coupled orbitals for each $m$, due to the time-consuming nature of the calculation.

For numerical calculations, we selected a set of parameters corresponding to $^{87}\mathrm{Rb}$ gas loaded in a bubble-shaped trap of radius $R=77\; \mu\mathrm{m}$ rotating at a frequency of $\Omega=20\; \hbox{Hz}$. The density is taken as $n=1.5\times 10^{12}\;\mathrm{cm}^{-3}$ \cite{Lundblad_npj_Microgravity2019,Tononi_PRL2020}, and the velocity of sound  can be estimated as $c\approx380\; \mu\mathrm{m}/\hbox{s}$. As a result, the natural frequency scale and the Tkachenko mode velocity are $m_{\mathrm{a}} c^2\approx200\; \mathrm{Hz}$ and $\cT\approx 60 \; \mu \mathrm{m}/s $. The intervortex distance and penetration length of the edge states are $l_{\mathrm{v}}\approx4.6\;\mu \mathrm{m}$ (around the poles) and  $l_{\mathrm{e}}\approx9.6\; \mu \mathrm{m}$, respectively. Notably, the hierarchy of spatial scales follows $l_{\mathrm{v}}\lesssim l_{\mathrm{e}}\ll R$, and we have approximately $10^3$ vortices/antivortices per hemisphere, justifying the adopted hydrodynamic framework.

\begin{figure}[t]
	\centering
	\includegraphics[width=\linewidth]{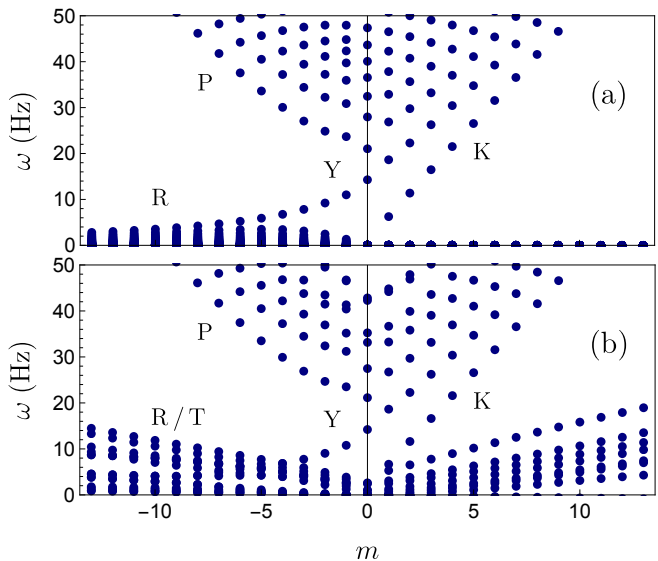}
	\caption{Spectrum of a rotating bubble-trapped (a) an inviscid classical fluid and (b) a superfluid. The presence of Tkachenko shear sound modes, supported by the Abrikosov lattice embedded in the superfluid, significantly impacts the low-frequency portion of the spectrum and limits the topological protection of the equatorial Kelvin and Yanai waves.}
	\label{fig:sphere_dispersion}
	\vspace{-0.2in}
\end{figure}

In the case where the Abrikosov lattice is assumed to be frozen, the superfluid behaves like a classical, inviscid fluid. As shown in Fig.~\ref{fig:sphere_dispersion}(a), the spectrum for this system in a spherical geometry includes dispersion curves for chiral Kelvin and Yanai modes that only propagate in one direction along the \emph{equator} of the trap. In the frequency range well above the gap, $2|\Omega|$, the effect of the rotation is insignificant, so the frequency modes can be approximated by the hydrodynamic excitation frequencies for the non-rotating spherical superfluid $\omega_l=c \sqrt{l(l+1)/R^2}$, where $l$ is the orbital quantum number. Rossby modes, which are influenced by the Coriolis force, also exist at low frequencies but are topologically trivial and can propagate in both directions. The high-frequency portion of the spectrum also includes Poincar\'{e} modes. For the regime considered in this case ($l_{\mathrm{e}} \ll R$), the lowest-frequency mode with $m=0$ can be estimated as $\omega_\mathrm{R}\approx 2|\Omega| \sqrt{3 l_{\mathrm{e}}/ R}$ and starts at a frequency well below the threshold of $2|\Omega|$. 

The spectrum for a rotating bubble-trapped superfluid is shown in Fig.~\ref{fig:sphere_dispersion}(b). The high-frequency region, which includes Poincar\'{e}, Kelvin, and Yanai modes, is only slightly modified compared to the spectrum of an inviscid classical fluid shown in Fig.~\ref{fig:sphere_dispersion}(a). Furthermore, not all corresponding eigenfrequencies have an imaginary part, making these modes stable. However, the presence of Tkachenko shear sound modes supported by the Abrikosov lattice significantly affects the low-frequency part of the spectrum. Firstly, the Rossby modes overlap with the Tkachenko modes, and as shown in the Appendix, these modes have a nonzero imaginary part and therefore become unstable. Secondly, chiral Kelvin and Yanai modes coexist with the Tkachenko waves and do not have complete topological protection. However, if the rotational symmetry is respected (e.g., if the superfluid density can be approximated as uniform), the coupling between these modes is inefficient due to the large momentum transfer required.

The edge mode behavior described here is generally applicable and only weakly dependent on $R$. Within our numerical calculation capabilities, the range $R/l_\mathrm{e}\sim 7-20$ is optimal for observing equatorial modes. For smaller bubble-shaped traps with $R\sim l_\mathrm{e}$, the Kelvin and Yanai equatorial modes are affected by finite-size quantization effects and are pushed out of the frequency gap. As the radius of the trap $R$ increases, the frequency range separating hybrid Tkachenko/Rossby waves and Poincar\'{e} waves from chiral Kevin and Yanai modes gradually shrinks (as $\omega_\mathrm{R}\approx 2|\Omega| \sqrt{3 l_{\mathrm{e}}/ R}$). 

\section{Discussion}

Chiral edge modes, as discussed, can be probed in experiments using two-photon Bragg scattering \cite{Stenger_PRK_1999,Stamper_PRL_1999,Vogels_PRL_2002}, which involves shining two laser beams intersecting at a finite angle onto the condensate to generate a moving optical lattice that can scatter with underlying elementary excitations. The resulting effects can be measured using the dynamic structure factor \cite{Ozeri_RMP_2005}.
In the case of a shell-shaped BEC, if the radius $R$ is large enough, the local behavior will be similar to that of a planar BEC, allowing for the use of Bragg spectroscopy to focus on chiral excitations along the equator of the bubble. Although the concentration of the Abrikosov vortex lattice is lower around the equator in comparison to that around the poles, equatorial modes do interact with vibrations of the vortex lattice (Tkachenko sound waves) in the upper and lower hemispheres, and their stability is not generic. Our calculations show that there is a range of parameters (trap radius and rotation frequency) where the equatorial Kelvin and Yanai modes (but not the Rossby ones) are very stable.

At finite temperatures, some atoms can escape the BEC and contribute to the normal fluid component, which has been neglected until now. As shown in the Appendix, two-fluid hydrodynamic equations that include the normal fluid can also be reformulated as an eigenvalue problem and topologically classified. As a result, we expect the presence of another set of equatorial waves that correspond to the second sound mode and are coupled to temperature–entropy oscillations propagating along the trap equator. These waves should be detectable in experiments.

In conclusion, the dynamics of a rotating bubble-trapped superfluid is more complicated than a inviscid classical fluid because of the accompanying Abrikosov vortex lattice. Contrary to the idea that such a system cannot host stable edge modes due to the lack of a global gap, we proved that partial or full topological protection for the chiral Kelvin and Yanai edge modes is possible if one considers both the compressional and shear stress of the vortex lattice in the dynamical equations. In contrast, the Rossby modes that are known to play an important role in the Earth climate dynamics become unstable, since they effectively mix with the Tkachenko vibrational modes of the vortex lattice. The predictions of our theory can be readily tested in the Gross-Pitaevskii equation simulations~\cite{SaitoArxiv2022} and in the state-of-the-art ground-based flat geometry~\cite{CoddingtonPRL2003} and bubble-trapped microgravity~\cite{CarolloNature2022} experiments.

\acknowledgments{We acknowledge fruitful discussions
with Tapio Samula, Matt Davis, and Eli Estrecho and support from the Australian Research Council Centre of Excellence in
Future Low-Energy Electronics Technologies.}

\begin{appendix}

\section{Hydrodynamic framework at finite temperature}\label{sec:Hamiltonian_finite_T}

In the main text, we assumed that the bubble-trapped superfluid was at low temperature and therefore did not include the normal fluid component in our analysis. However, in this section, we extend our hydrodynamic framework to include the normal fluid dynamics as well. The two-fluid hydrodynamic equations are  \cite{SoninRMP1987,BookPitaevskii}:
\begin{align}
   &\partial_t \rho +\rho \nabla \cdot \V = 0, \\
   &\partial_t \s +\nabla \cdot(\s\v_n)=0, \\   
   &\partial_t \v_s= -2 \O\times\v_s-\nabla \mu+\f,\\
   &\partial_t \V= -2\O\times\V- \frac{1}{\rho}\nabla P+\frac{\rho_s}{\rho}\f=0,\\
    &2\O\times(\partial_t\e-\v_s)=-\f,\\ 
    & \f=\cT^2\left[\nabla^2\e-2\alpha \nabla(\nabla\cdot\e) \right].
\end{align}
Here $\rho=\rho_s+\rho_n$ is the total density with $\rho_{s(n)}$ is the density of superfluid (normal) component and $\s$ is the entropy density. The velocity of the center-of-mass and the relative motion are $\V=(\rho_s\v_s+\rho_n\v_n)/\rho$ and $\u=\v_n-\v_s$, where $\v_{s(n)}$ is the velocity of the corresponding fluid component.
We choose two independent thermodynamic functions as $T$ and $P$ and thermodynamic variables as $\rho$ and $\st=\s/\rho$ \cite{BookPitaevskii}. The chemical potential can be eliminated by using the relation $\nabla\mu=\nabla P/\rho-\s\nabla T /\rho$ \cite{SoninRMP1987}.
%%%%%%%%%% 
% the subscript convention I followed Pitaevskii's book, i.e. no \mathrm
%%%%%%%%%%

Similar to the main text, we rewrite the above equations by the chiral basis as follows:
\begin{align}
    J_{V\pm}&=\frac{\rho_0(V_x\pm i V_y)}{\sqrt{2}}, &J_{V0}&=c_1 \rho,\\
    J_{u\pm}&=\rho_0\sqrt{\frac{\rho_s \rho_n}{\rho^2}}\frac{u_x\pm i u_y}{\sqrt{2}}, &J_{u0}&=c_2 \frac{\rho_0^2}{\rho_s},\\
    J_{\epsilon\pm}&=\rho_0 \sqrt{\frac{\rho_s}{\rho}} \cT q \frac{\epsilon_x\pm i\epsilon_y}{\sqrt{2}}.
\end{align}
With the spinor defined as:
\begin{align}
    \psi=[\Jepsp, J_{V+}, J_{V0}, J_{V-}, J_{u+}, J_{u0}, J_{u-}, \Jepsm]^\mathsf{T},
\end{align}
we arrive at the $8$ by $8$ Hamiltonian:
\begin{widetext}
\begin{equation}\label{eq:H8}
\hat{\mathcal{H}}_T=\begin{bmatrix}
    \bar{\alpha} \frac{\cT^2}{2\Omega} q^2     &   i\sqrt{\frac{\rho_s}{\rho}}\cT q   &  0  &  0  &  -i\sqrt{\frac{\rho_n}{\rho}} \cT q  &  0  &  0  &  -\alpha \frac{\cT^2}{2\Omega}q_+^2\\
-i\bar{\alpha}\sqrt{\frac{\rho_s}{\rho}}\cT q  &             2\Omega               &   \frac{c_1 q_+}{\sqrt{2}}  & 0 & 0 & 0 & 0 & i \alpha \sqrt{\frac{\rho_s}{\rho}}\cT q_+^2\\
    0                                       &   \frac{c_1 q_-}{\sqrt{2}} &   0  &  \frac{c_1 q_+}{\sqrt{2}}& 0 & 0 & 0 & 0 \\
i\alpha\sqrt{\frac{\rho_s}{\rho}}\cT \frac{q^2_-}{q}  &  0 & \frac{c_1 q_-}{\sqrt{2}}  & -2\Omega & 0 & 0 & 0 & -i \bar{\alpha}\sqrt{\frac{\rho_s}{\rho}}\cT q\\
i\bar{\alpha}\sqrt{\frac{\rho_n}{\rho}}\cT q & 0 & 0 & 0 & 2\Omega & \frac{c_2 q_+}{\sqrt{2}}   & 0 & -i\alpha\sqrt{\frac{\rho_n}{\rho}}\cT \frac{q^2_+}{q} \\
0 & 0 & 0 & 0 & \frac{c_2 q_-}{\sqrt{2}}  & 0 & \frac{c_2 q_+}{\sqrt{2}} & 0\\
-i\alpha \sqrt{\frac{\rho_n}{\rho}}\cT \frac{q_-^2}{q} & 0 & 0 & 0 & 0 & \frac{c_2 q_-}{\sqrt{2}}  & -2\Omega & i\bar{\alpha} \sqrt{\frac{\rho_n}{\rho}}\cT q\\
\alpha \frac{\cT^2}{2\Omega} q_-^2  & 0 & 0 & i\sqrt{\frac{\rho_s}{\rho}}\cT q & 0 & 0 & -i \sqrt{\frac{\rho_n}{\rho}}\cT q & - \bar{\alpha}\frac{\cT^2}{2\Omega} q^2
    \end{bmatrix},
\end{equation}
\end{widetext}
where $\bar{\alpha}=1-\alpha$ and $q_{\pm}=q_x\pm i q_y$; $c_1$ and $c_2$ are the first and second sound respectively \cite{BookPitaevskii}. They are defined as:
\begin{equation}
    c_1^2=\left(\frac{\delta P}{\delta \rho}\right)_{\st} \quad\quad\text{and}\quad\quad c_2^2=\frac{\rho_s}{\rho_n} \st^2 \left(\frac{\delta T}{\delta \st} \right)_\rho .
\end{equation}
We have neglected the thermodynamic derivatives $(\delta T/\delta\rho)_{\st}$ and $(\delta P/\delta\st)_{\rho}$ as they describe the interaction between the center-of-mass and relative motion. This approximation is valid when the first- and second-sound modes do not interact or hybridize. 

\begin{figure}[h]
	\centering
\includegraphics[width=\linewidth]{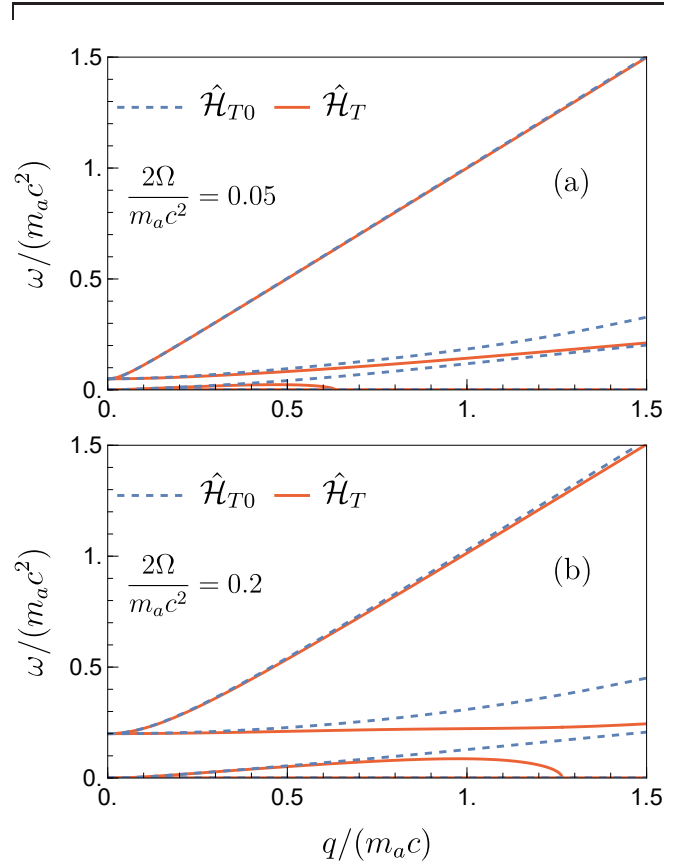}
	\caption{Dispersion relations for the finite temperature Hamiltonian $\hat{\mathcal{H}}_T(\q)$ (solid red line) and the approximated Hamiltonian $\hat{\mathcal{H}}_{T0}(\q)$ (blue dashed line) that neglects the compressional contribution to the elastic force density, with $\rho_s/\rho=0.5$, $c_1/c=4.2$, $c_2/c=0.6$ \cite{Hybridization_sound}.}
	\label{fig:E8_dispersion}
\end{figure}

In Fig.~\ref{fig:E8_dispersion}, we compare the spectra of $\hat{\mathcal{H}}_T$ its Hermitian version ($\alpha=0$) and the non-Hermitian one ($\alpha=1$). It can be seen that the Hermitian case represents the low momentum limit of the full non-Hermitian Hamiltonian. The spectrum of the full Hamiltonian maintains the global gap between the inertial modes and the Tkachenko mode which saturates at a maximum value. This gap could also accommodate a set of chiral edge modes, as mentioned in the main text. In the limit of the vanishing normal density, $\rho_n\to 0$, we arrive to the zero temperature Hamiltonian $\hat{\mathcal{H}}$ appearing in the main text. 

At finite temperature, according to \cite{GiffordPRA2008}, thermal fluctuations can lead to the Kosterlitz-Thouless transition (or BKT transition) of vortex lattice characterised by a melting temperature $T_m$. Below $T_m$ the vortex lattice is stable; and above $T_m$ it will melt. Monte Carlo simulation in \cite{KragsetPRL2006} showed that, depending on the specific parameters, it is possible that the ratio $T_m/T_c=0.76$, where $T_c$ is the usual BEC (or BKT) critical temperature. Hence, the vortex lattice can be stable in typical experimental conditions.

In fact, the superfluid equatorial modes will become more stable if the vortex lattice has melted, because the additional decay channel for them (decay into the Tkachenko waves) will cease to exist, see Fig.~3(a) in the main text.

The extension of the BKT transition to a spherical setting is highly nontrivial, which has been achieved only recently in \cite{TononiPRR2022} for non-rotating bubble-trapped BECs. So does the quantitative description of the transition temperature for a rotating bubble-trapped BEC, which merits additional study. Our analysis will be relevant as long as we are sitting below $T_m$ where the two-ﬂuid hydrodynamic description applies.

\section{Chern Number Calculation}\label{sec:Chern_Number_Calculation}

In this section, we demonstrate the calculation of the Chern number for the inertial modes. We start by examining the $3$ by $3$ Hermitian Hamiltonian $\hat{\mathcal{H}}^\prime_0(\vec{q})$:
\begin{equation}\label{eq:Hpr}
\hat{\mathcal{H}}^\prime_0(\q)=
    \begin{bmatrix}
2\Omega  &   \frac{c (q_x+iq_y)}{\sqrt{2}}  &  0  \\  \frac{c (q_x-i q_y)}{\sqrt{2}}  &   0   &      \frac{c (q_x+i q_y)}{\sqrt{2}} \\
    0   &   \frac{c (q_x-i q_y)}{\sqrt{2}}   &   -2\Omega    
    \end{bmatrix}=\bm{\sigma}\cdot\Q,
\end{equation}
where we have decomposed the matrix using the spin-$1$ Pauli matrix (with the reminder that $\hbar=1$):
\begin{align}
    \sigma_x&=\frac{1}{\sqrt{2}}\begin{bmatrix}
    0 & 1 & 0\\
    1 & 0 & 1\\
    0 & 1 & 0
    \end{bmatrix},\quad \sigma_y=\frac{1}{\sqrt{2}}\begin{bmatrix}
    0 & -i & 0\\
    i & 0 & -i\\
    0 & i & 0
    \end{bmatrix},\\[1em]
    \sigma_z&=\begin{bmatrix}
    1 & 0 & 0\\
    0 & 0 & 0\\
    0 & 0 & -1
    \end{bmatrix},\quad \text{with } \Q\equiv(c q_x, -\, c q_y, 2\Omega),
\end{align}
where the negative sign in $\Q$ in its $y$-component will not change any of our physical discussions because $\hat{\mathcal{H}}^\prime_0(\q)$ is Hermitian. There are standard techniques in calculating the Berry phase of such spinor system evolving along an external magnetic field, which is our $\Q$. The Berry phase is simply the flux through the area, bounded by the evolution path of $\Q$, of a momopole of strength $n_\mathcal{S}$ (the eigenvalue of the spin, $n_\mathcal{S}=-\mathcal{S}, -\mathcal{S}+1, \dots, \mathcal{S}$, where $\mathcal{S}$ is the total spin) located at the origin \cite{Book_TI}. Simply speaking, if we trace $\Q$ in the whole momentum plane (note that $\abs{\q}$ can be large enough so that $\Q$ is effectively rotating on a plane), then $\Q$ subtends a solid angle of $2\pi$. For a spin-$1$ system $n_\mathcal{S}=-1, 0, 1$, by dividing the Berry phase by $2\pi$ we get the Chern number as $-1$, $0$, and $+1$.

In the following, we demonstrate the explicit calculation of the Chern number without relying on the general results mentioned above.
We define $R_Q\equiv\abs{\Q}=\sqrt{c^2q^2+4\Omega^2}$ and the parameterization $\sin{\vartheta}\equiv c q/R_Q$ and $\cos{\vartheta}\equiv 2\Omega/R_Q$, which is based on the relation: $(c q/R_Q)^2+(2\Omega/R_Q)^2=1$.
The eigenvalues of $\hat{\mathcal{H}}^\prime_0(\q)$ are $-R_Q$, $0$ and $R_Q$ and the corresponding wave functions are given by
\begin{equation}
\label{eq:v03}
\begin{split}
\v_{01}&=\{ \sin^2\frac{\vartheta}{2}\, e^{i \phi}, -\sqrt{2}\sin\frac{\vartheta}{2}\cos\frac{\vartheta}{2}, \cos^2\frac{\vartheta}{2}\,e^{-i \phi}\}, \\
\v_{02}&=\{-\frac{1}{\sqrt{2}} \sin\vartheta e^{i\phi}, \cos\vartheta, \frac{1}{\sqrt{2}}\sin\vartheta e^{-i\phi} \} , \\
\v_{03}&=\{\cos^2\frac{\vartheta}{2}\,e^{i \phi}, 
    \sqrt{2}\sin\frac{\vartheta}{2}\cos\frac{\vartheta}{2}, 
    \sin^2\frac{\vartheta}{2}\,e^{-i \phi}
    \}.
\end{split}
\end{equation}
where $\phi$ is the polar angle for $\q$. Next we calculate the Chern number for the upper inertial mode $E_{03}^\prime$ and $\v_{03}$.
The Berry connection is given by:
\begin{equation*}\label{eq:Berry_connection}
    \A=\ev{i\nabla_\q}{\v_{03}},\quad\text{where }\nabla_\q=\mathbf{e}_q \partial_q +\mathbf{e}_\phi (1/q)\partial_\phi.
\end{equation*}
Inserting Eq.~\eqref{eq:v03} into the expression of $\A$, we have the radial and azimuthal components: $A_q=0$ and $A_\phi=-\cos\vartheta/ q$. The Berry curvature is given by:
\begin{equation}
    \mathbf{B}=\nabla_\q\times\A=\ez \frac{1}{q}\partial_q(q A_\phi)=-\ez \frac{1}{q}\partial_q\cos\vartheta.
\end{equation}
Then the Chern number (in units of $2\pi$) is given by the integration of the Berry curvature over the whole momentum plane:
\begin{align*}
    \mathcal{C}_{03}&=\frac{1}{2\pi}\int_0^\infty \diff q \int_0^{2\pi} \diff\phi\,q\, \mathbf{B}\cdot \ez
    =-\int_0^\infty \diff q \,\partial_q\cos\vartheta\\
    &= \cos\vartheta\bigg|^{q=0}_{q\to\infty}
    =\frac{2\Omega}{\sqrt{c^2q^2+4\Omega^2}}\bigg|^{q=0}_{q\to\infty}\nn
    =1. \nn
\end{align*}
The Chern number for the other modes can be calculated in a similar manner, resulting in $\mathcal{C}_{01}=-1$ and $\mathcal{C}_{02}=0$. For the $5$ by $5$ Hermitian Hamiltonian $\hat{\mathcal{H}}_0$, we can use the same procedure to determine the Chern number for the inertial modes as $\mathcal{C}=\Omega/\abs{\Omega}$.

\section{Elastic Stress Tensor of Vortex Lattice on a Sphere}\label{sec:Stress_Tensor}

In this section, we derive the hydrodynamic equations for a system on a sphere. The central step in this process is obtaining the force density due to the shear and compressional stress of the vortex lattice. We will first review the derivation of this expression in a planar system, following Ref.~\cite{Baym1986}, and then extend it to a spherical surface using some principles of differential geometry.

We use symbols with one free index to represent contravariant vectors, and those with two free indexes to represent second-order tensors, which can also be written as matrices. When an index is repeated, it implies summation.

The vortex elastic stress tensor is given by \cite{Baym1986}:
\begin{equation}
    \gamma_{ik}=\frac{\Omega}{4 m_a}(\partial_k \epsilon_i+\partial_i \epsilon_k-3\delta_{ik}\partial_j \epsilon_j),
\end{equation}
where $i,j,k=x,y$ and $\delta_{ik}$ is the Kronecker delta. Written explicitly, each component reads:
\begin{equation}
    \gamma_{ik}=\frac{\Omega}{4 m_a}\begin{bmatrix}
    -\partial_x\epsilon_x-3\partial_y \epsilon_y & \partial_y\epsilon_x+\partial_x\epsilon_y\\
    \partial_y\epsilon_x+\partial_x\epsilon_y & -3\partial_x\epsilon_x-\partial_y\epsilon_y
    \end{bmatrix}
\end{equation}
The vortex elastic force density is defined as $\tilde{f}_i=-\partial_k \gamma_{ik}$ and is given by  
\begin{align}
\tilde{f}_x&=\frac{\Omega}{4 m_a}\left(\partial_x^2\epsilon_x+2\partial_x\partial_y\epsilon_y-\partial_y^2\epsilon_x \right),\\
\tilde{f}_y&=\frac{\Omega}{4 m_a}\left(-\partial_x^2\epsilon_y+2\partial_x\partial_y\epsilon_x+\partial_y^2\epsilon_y \right). 
\end{align}
It gives the expression of 
\begin{equation}
\tilde{\mathbf{f}}= \frac{\Omega}{4 m_a} \left[2\nabla(\nabla\cdot\e)-\nabla^2\e      \right].
\end{equation}

To extend the hydrodynamic equations to a spherical surface, the standard method is to replace the spacial partial derivatives with covariant derivatives along the longitude and latitude directions, such that $\partial_x\to\nabla_\theta$ and $\partial_y\to\nabla_\varphi$ \cite{Book_docarmo}.
The stress tensor in the $(\theta,\varphi)$ parametrization reads:
\begin{equation}
    \gamma_{ik}=\frac{\Omega_0 \abs{\cos\theta}}{4 m_a}\begin{bmatrix}
    -\nabla_\theta\epsilon_\theta-3\nabla_\varphi \epsilon_\varphi & \nabla_\varphi\epsilon_\theta+\nabla_\theta\epsilon_\varphi\\
    \nabla_\varphi\epsilon_\theta+\nabla_\theta\epsilon_\varphi & -3\nabla_\theta\epsilon_\theta-\nabla_\varphi\epsilon_\varphi
    \end{bmatrix},
\end{equation}
where we have included the $\theta$-dependent local rotation speed in the expression. By using the relation $\tilde{f}_i=-\nabla_k \gamma_{ik}$, we can then obtain the elastic force density as follows:
\begin{equation}
    \tilde{\mathbf{f}}=\frac{\Omega_0\abs{\cos\theta}}{4 m_a} \left[2\nabla (\nabla\cdot\e)-\nabla^2\e      \right]+\operatorname{F}[\sin\theta],
\end{equation}
where the absolute value of $\cos\theta$ ensures that the local Tkachenko mode velocity $\cT=\sqrt{\abs{\Omega_0 \cos\theta}/4 m_a}$ is positive in both hemispheres, and the $\nabla$ operator takes the gradient along the tangent direction on the sphere.
The last term $\operatorname{F}[\sin\theta]$ represents all $\sin\theta$-dependent terms that arise from $\nabla_\theta(\cos\theta)$ when applying the Leibniz rule for the covariant derivative operator. As the local rotation speed approaches $0$ near the equator, the lattice constant of the vortex lattice becomes larger and larger, suppressing all other length scales in the system. In this case, the hydrodynamic description becomes unrealistic as the coarsed-grained continuous model approximation breaks down. Therefore, the effects of $\operatorname{F}[\sin\theta]$ will only appear beyond the validity regime of the hydrodynamic model. For the purpose of demonstrating the equatorial modes, it is sufficient to keep only the first term and neglect $\operatorname{F}[\sin\theta]$.
The hydrodynamic equations on a sphere are then as follows:
\begin{align}
   & \partial_t \rho +\rho_0\nabla \cdot \v = 0,\label{eq:hydro_sphere_1}\\
    &\partial_t \v= -2 \Omega_0 \cos\theta\,\er\times\v-\frac{c^2}{\rho_0}\nabla \rho +\f, \\
   & 2\Omega_0\cos\theta\,\er\times\left(\partial_t\e-\v\right)=-\f,\\
   &\f=\frac{\Omega_0\abs{\cos\theta}}{4 m_a} \left[\nabla^2 \e  -\alpha\,2\nabla(\nabla\cdot\e)  \right], \label{eq:hydro_sphere_2}
\end{align}
where $\er$ is the unit vector in the radial direction. Note that we have adjusted the sign of the definition of $\mathbf{f}$ in Eq.~\eqref{eq:hydro_sphere_2} to be consistent with the convention used in Refs.~\cite{BaymPRL2003,SoninPRA2005}.

\section{Vector Spherical Harmonics Expansion}\label{sec:vector_Spherical_Harmonics}

To solve Eq.\eqref{eq:hydro_sphere_1}-\eqref{eq:hydro_sphere_2},
we expand the dynamical functions $\rho$, $\v$, and $\e$ using a set of orthonormal basis called vector spherical harmonics (VSH) \cite{VSHwiki,CooperPRB2022}, which are defined through the spherical harmonics $Y_{lm}(\theta,\varphi)$ \cite{Book_Jackson} as follows:
\begin{align}
    &\Ylm=\frac{Y_{lm} \er}{\sqrt{l(l+1)}},\,
    \Psilm=\frac{r\nabla Y_{lm}}{\sqrt{l(l+1)}},\,
    \Philm=\frac{r\, \er\times\nabla Y_{lm}}{\sqrt{l(l+1)}},\\
    &l =1, 2, 3, \dots \text{ and } m =-l, -l+1, \dots, l \nn
\end{align}
Additionally $\mathbf{Y}_{00}=\er \sqrt{1/4\pi}$ and $\mathbf{\Psi}_{00}=\mathbf{\Phi}_{00}=\mathbf{0}$ and thus the $l=m=0$ expansion coefficient will be arbitrary.
The expansion reads:
\begin{align}
    \v&=\sum_\lm (v^\Psi_\lm \Psilm+v^\Phi_\lm \Philm ),\quad\rho=\sum_\lm\rho_\lm Y_\lm, \label{eq:VSH_expansion}\\
    \e&=\sum_\lm (\epsilon^\Psi_\lm \Psilm+\epsilon^\Phi_\lm \Philm).\label{eq:VSH_expansion2}
\end{align}
By inserting Eq.~\eqref{eq:VSH_expansion}-\eqref{eq:VSH_expansion2} into Eq.~\eqref{eq:hydro_sphere_1}-\eqref{eq:hydro_sphere_2} and then use the orthogonality relation of $\Psilm,\Philm$ by applying $\int\diff S\,\Psilm^*\cdot(\bullet) $ and $\int\diff S\,\Philm^*\cdot(\bullet) $, where $\int\diff S = \int_0^{2\pi}\diff\varphi\int_0^\pi\sin\theta\diff\theta$ represents the integration over the solid angle, we have
\begin{equation}
\quad\partial_t\rho_\lm=\rho_0 q_l v_\lm^\Psi,\label{eq:VSH_Cartesian_1}
\end{equation}
accompanied with
\begin{widetext}
\begin{align}
& \begin{cases}
         & \partial_t v_\lm^\Psi+2\Omega_0\sum_{l^\prime}\left[(\partial_t\epsilon_{l^\prime m}^\Psi)M_{nn^\prime}^{\Psi\Phi}-(\partial_t\epsilon_{l^\prime m}^\Phi)M_{nn^\prime}^{\Psi\Psi}   \right]=-\frac{c^2 q_l}{\rho_0}\rho_\lm , \\[0.5em]
         & \partial_t v_\lm^\Phi+2\Omega_0\sum_{l^\prime}\left[(\partial_t\epsilon_{l^\prime m}^\Psi)M_{nn^\prime}^{\Phi\Phi}-(\partial_t\epsilon_{l^\prime m}^\Phi)M_{nn^\prime}^{\Phi\Psi}   \right]=0,
    \end{cases}   \label{eq:VSH_Cartesian_2} \\[0.5em]
    & \begin{cases}
    &\sum_{l^\prime}\left[(\partial_t \epsilon_{l^\prime m}^\Psi-v_{l^\prime m}^\Psi)M_{n n^\prime}^{\Psi\Psi}+ (\partial_t \epsilon_{l^\prime m}^\Phi-v_{l^\prime m}^\Phi)M_{n n^\prime}^{\Psi\Phi}    \right]=\sum_{l^\prime}\left[-\epsilon_{l^\prime m}^\Psi \omega'_{l'}W_{n n^\prime}^{\Psi\Psi}+\epsilon_{l^\prime m}^\Phi \omega'_{l'} W_{n n^\prime}^{\Psi\Phi}   \right],\\[0.5em]
    &\sum_{l^\prime}\left[(\partial_t \epsilon_{l^\prime m}^\Psi-v_{l^\prime m}^\Psi)M_{n n^\prime}^{\Phi\Psi}+ (\partial_t \epsilon_{l^\prime m}^\Phi-v_{l^\prime m}^\Phi)M_{n n^\prime}^{\Phi\Phi}    \right]=\sum_{l^\prime}\left[-\epsilon_{l^\prime m}^\Psi \omega'_{l'} W_{n n^\prime}^{\Phi\Psi}+\epsilon_{l^\prime m}^\Phi \omega'_{l'} W_{n n^\prime}^{\Phi\Phi}   \right],
    \end{cases} \label{eq:VSH_Cartesian_3}
\end{align}
\end{widetext}
where in some subscripts the symbol $n=l m$ and $n^\prime=l^\prime m$ are used as shorthand notations. Here we have also introduced $q_l=\sqrt{l(l+1)}/R$ and $\omega'_l=l(l+1)/4 m_a R^2$. The $\int_0^{2\pi}\diff\varphi$ integration gives an overall $\delta_{m,m^\prime}$ signaling the $z$-axis symmetry such that the dynamics do not couple different $m$ components.
The $M$ and $W$ matrices stem from the expansion of the $\cos\theta$ and $\abs{\cos\theta}$ respectively \cite{CooperPRB2022}:
\begin{widetext}
\begin{align}
    M_{n n^\prime}^{\Psi \Psi} &= \int\diff S\, \cos\theta \mathbf{\Psi}_{n}^* \cdot \mathbf{\Psi}_{n^\prime}=\delta_{m m^\prime}\left(\delta_{l,l^\prime-1}\mu^S_{l m} +\delta_{l,l^\prime+1}\mu^S_{l^\prime m} \right)&\qquad
    M_{n n^\prime}^{\Psi \Phi} &= \int\diff S\, \cos\theta \mathbf{\Psi}_{n}^* \cdot \mathbf{\Phi}_{n^\prime}=i\delta_{n n^\prime}\mu^A_{lm}\\
    M_{n n^\prime}^{\Phi \Phi} &= \int\diff S\, \cos\theta \mathbf{\Phi}_{n}^* \cdot \mathbf{\Phi}_{n^\prime}=\delta_{m m^\prime}\left(\delta_{l,l^\prime-1}\mu^S_{l m} +\delta_{l,l^\prime+1}\mu^S_{l^\prime m} \right) &\qquad
    M_{n n^\prime}^{\Phi \Psi} &= \int\diff S\, \cos\theta \mathbf{\Phi}_{n}^* \cdot \mathbf{\Psi}_{n^\prime}=-i\delta_{n n^\prime}\mu^A_{lm}
\end{align}
\end{widetext}
The integral in $M$ can be evaluated analytically by using properties of the associated Legendre functions $P_l^m$ (see  Ref.~\cite{Book_integrals} for more information). Only nearby modes with $l$ difference of $\pm 1$ will give non-zero results:
\begin{align}
    \mu^S_{l m}&=\frac{\sqrt{(l+1)^2-m^2}}{l+1}\sqrt{\frac{l(l+2)}{(2l+1)(2l+3)}},\\ \mu^A_{l m}&=-\frac{m}{l(l+1)}\qquad \text{with } l\neq0
\end{align}
If $l=0$, then $\mu^A_{0 0}=0$. The superscripts $S$ and $A$ stand for symmetric and asymmetric terms respectively.

The $W$ matrix is defined as expansion of $\abs{\cos\theta}$ as:
\begin{align}
    W_{n n^\prime}^{\Psi \Psi} &= \int\diff S\, \abs{\cos\theta} \mathbf{\Psi}_{n}^* \cdot \mathbf{\Psi}_{n^\prime}=\mu^C_{nn^\prime} \\
    W_{n n^\prime}^{\Psi \Phi} &= \int\diff S\, \abs{\cos\theta} \mathbf{\Psi}_{n}^* \cdot \mathbf{\Phi}_{n^\prime}=i\mu^B_{nn^\prime}\\
    W_{n n^\prime}^{\Phi \Phi} &= \int\diff S\, \abs{\cos\theta} \mathbf{\Phi}_{n}^* \cdot \mathbf{\Phi}_{n^\prime}=\mu^C_{nn^\prime} \\
    W_{n n^\prime}^{\Phi \Psi} &= \int\diff S\, \abs{\cos\theta} \mathbf{\Phi}_{n}^* \cdot \mathbf{\Psi}_{n^\prime}=-i\mu^B_{nn^\prime}
\end{align}
The integral admits analytic expression given any specific value of $l$ and $m$.

We can rewrite Eq.~\eqref{eq:VSH_Cartesian_1}-\eqref{eq:VSH_Cartesian_3} in the chiral basis:
\begin{align}
    J^{\pm}_\lm&=(v^\Psi_\lm\pm i v^\Phi_\lm)/\sqrt{2},\quad J^0_\lm=c \rho_\lm/ \rho_0,\\ \quad
    & \quad \; \quad  K^{\pm}_\lm=(\epsilon^\Psi_\lm\pm i \epsilon^\Phi_\lm)/\sqrt{2}.
\end{align}
Note that $K^{\pm}_\lm$ is seemingly having a different dimension compared with $J^0_\lm$ and $J^\pm_\lm$, nevertheless,  the transformed equation will have the correct dimensions in both the left- and right-hand sides. The transformed Eq.~\eqref{eq:VSH_Cartesian_1}-\eqref{eq:VSH_Cartesian_3} leads to 
\begin{equation}\label{eq:J0lm}
\omega J^0_\lm = i\frac{c q_l}{\sqrt{2}} (J^+_\lm +  J^-_\lm) 
\end{equation}
accompanined by 
\begin{equation}\label{eq:Klm}
\begin{split}
\omega \mu^A_\lm K^\pm_\lm = \mp \omega\mu^S_\lm K^\pm_{l+1,m} \mp \omega\mu^S_{l-1,m} K^\pm_{l-1,m} \\
+ i \mu^A_\lm J^\pm_\lm  \pm i \mu^S_\lm J^\pm_{l+1,m} \pm i \mu^S_{l-1,m} J^\pm_{l-1,m} \\
- \sum_{l^\prime=m}^\infty \omega'_{l'}(  \mu^B_{l l^\prime m} \pm \mu^C_{l l^\prime m} )K^\mp_{l^\prime m},
\end{split}
\end{equation}
and
\begin{equation}\label{eq:Jlm}
\begin{split}
\omega J^\pm_\lm = -i\frac{c q_l}{\sqrt{2}} J^0_\lm + 2\Omega_0 \mu^A_\lm J^\pm_\lm \\ \pm 2\Omega_0 \mu^S_\lm J^\pm_{l+1,m} 
\pm 2\Omega_0 \mu^S_{l-1,m} J^\pm_{l-1,m} \\ +   \sum_{l^\prime=m}^\infty i \omega'_{l'}  2\Omega_0(\mu^B_{l l^\prime m}\pm\mu^C_{l l^\prime m})K^\mp_{l^\prime m}
\end{split}
\end{equation}
Note that the above equations \eqref{eq:J0lm}-\eqref{eq:Jlm}, when written in matrix form, do not represent an eigenvalue problem as the matrix on the left is degenerate and $\omega$ is scattered across the right-hand side of the equation. To determine $\omega$, we need to move all matrices to the left and set the determinant of the combined matrix to zero, with $\omega$ being a variable. The roots of this resulting polynomial equation correspond to the dispersion figures depicted in the main text.

\section{Complex Spectrum on a Spherical Surface}

\begin{figure}[h]
	\centering
	\includegraphics[width=\linewidth]{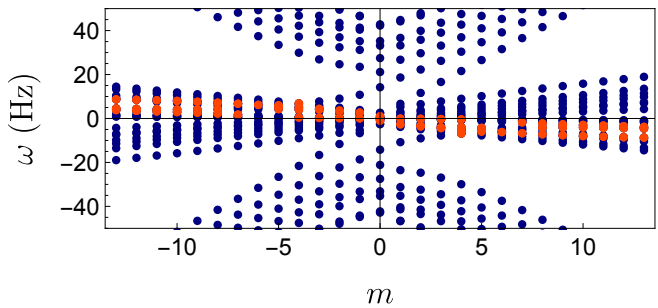}
	\caption{The spectrum of the hydrodynamic Hamiltonian on a sphere is shown by the blue dots, with $2\Omega_0/m_a c^2=0.2$ and $R/l_c=8$. Each red dot represents a pair of conjugate data points with positive and negative imaginary parts.}
	\label{fig:SM_sphere_dispersion_complex}
\end{figure}

The Hamiltonian $\hat{\mathcal{H}}$ is non-Hermitian, so its eigenvalues may be complex in general. However, on the surface of a sphere, the spectrum obtained is predominantly real, as shown in Fig.~\ref{fig:SM_sphere_dispersion_complex}, including the edge modes of interest. This indicates that these modes are stable, although they are not fully protected by a true energy gap. As shown in Fig.~\ref{fig:SM_sphere_dispersion_complex}, the edge modes are well protected for small values of $m$, meaning that scattering with the Tkachenko bulk is not very efficient because a large momentum exchange would be required.

%\end{widetext}
\end{appendix}

\bibliography{superfluid_Eq_wave_ref}

\end{document}